\documentclass[journal=esthag,manuscript=article]{achemso}

\usepackage{amsmath}
\usepackage{graphicx}
\usepackage{subcaption}
\usepackage{float}
\usepackage{lineno}
\usepackage{pdfpages}

\usepackage[sort&compress,numbers,super]{natbib}
\setkeys{acs}{doi = true}
\setkeys{acs}{maxauthors = 0}

\usepackage[version=3]{mhchem} % Formula subscripts using \ce{}

\usepackage{etoolbox} % needed for patch
\makeatletter
\patchcmd{\acs@contact@details}{E}{*\,E}{}{}
\patchcmd{\acs@email@list@aux}{;}{\par*\,Email}{}{}
\makeatother

\author{Ellen M. Considine}
\email{ellen_considine@g.harvard.edu}
\author{Jiayuan Hao}
\affiliation{Department of Biostatistics, Harvard T.H. Chan School of Public Health, Boston, Massachusetts, 02115, USA.}
\author{Priyanka deSouza}
\affiliation{Department of Urban and Regional Planning, University of Colorado Denver, University of Colorado Denver, Denver, Colorado, 80202, USA.}
\alsoaffiliation{CU Population Center, University of Colorado Boulder, Boulder, Colorado, 80309, USA.}
\author{Danielle Braun}
\affiliation{Department of Biostatistics, Harvard T.H. Chan School of Public Health, Boston, Massachusetts, 02115, USA.}
\alsoaffiliation{Department of Data Science, Dana-Farber Cancer Institute, Boston, Massachusetts, 02215, USA.}
\author{Colleen E. Reid}
\affiliation{Department of Geography, University of Colorado Boulder, Boulder, Colorado, 80302, USA.}
\alsoaffiliation{CU Population Center, University of Colorado Boulder, Boulder, Colorado, 80309, USA.}
\alsoaffiliation{Earth Lab, University of Colorado Boulder, Boulder, Colorado, 80303, USA.}
\author{Rachel C. Nethery}
\affiliation{Department of Biostatistics, Harvard T.H. Chan School of Public Health, Boston, Massachusetts, 02115, USA.}

\title{Evaluation of Model-Based PM$_{2.5}$ Estimates for Exposure Assessment During Wildfire Smoke Episodes in the Western U.S.}

 % couldn't get this to show up anywhere?

\begin{document}


Synopsis: Airborne particulate matter data spanning years and large areas are needed to study the health impacts of wildfire smoke. We examine how well existing large datasets capture smoke exposures.

\begin{abstract}
Investigating the health impacts of wildfire smoke requires data on people’s exposure to fine particulate matter (PM$_{2.5}$) across space and time. In recent years, it has become common to use machine learning models to fill gaps in monitoring data. However, it remains unclear how well these models are able to capture spikes in PM$_{2.5}$ during and across wildfire events. Here, we evaluate the accuracy of two sets of high-coverage and high-resolution machine learning-derived PM$_{2.5}$ data sets created by Di et al. (2021) and Reid et al. (2021). In general, the Reid estimates are more accurate than the Di estimates when compared to independent validation data from mobile smoke monitors deployed by the US Forest Service. However, both models tend to severely under-predict PM$_{2.5}$ on high-pollution days. Our findings complement other recent studies calling for increased air pollution monitoring in the western US and support the inclusion of wildfire-specific monitoring observations and predictor variables in model-based estimates of PM$_{2.5}$. Lastly, we call for more rigorous error quantification of machine-learning derived exposure data sets, with special attention to extreme events.
\end{abstract}

\section{Introduction}
\subsection{Motivation: Health Impacts of Wildfire Smoke}
Wildfires increasing PM$_{2.5}$ (fine particulate matter) in the United States (US) could undermine the strides the US has made in reducing non-smoke related PM$_{2.5}$ in recent decades \cite{burke_2021, odell_2019}. There is robust evidence of the severe health consequences of wildfire-related PM$_{2.5}$ exposure, ranging from premature mortality to asthma exacerbation \cite{neumann_2021, liu_2016, reid_maestas_2019}. However, there is mixed evidence for impacts of wildfire smoke on some health endpoints, such as cardiovascular disease \cite{chen_2021, reid_maestas_2019, reid_2016}. Many have argued that addressing this inconsistency in findings for the effects of wildfire smoke on some health endpoints, in addition to being able to study cumulative health impacts of wildfire smoke, requires smoke exposure assessment over large time periods and regions, enabling investigation across fire events. 

While wildfire smoke often exposes people to a large collection of harmful air pollutants (including inorganic gases such as carbon monoxide, ozone, and NO$_X$; hydrocarbons, and more\cite{naeher_2007}), the evidence to date points to PM$_{2.5}$ as the pollutant of  greatest concern for human health. Accordingly, epidemiologic studies evaluating the health impacts of wildfire smoke most commonly rely on estimates of ambient PM$_{2.5}$ concentrations for exposure assessment. Ground measurements of PM$_{2.5}$ from federal reference method (FRM) or federal equivalence method (FEM) monitors, used primarily for regulation of air pollution from point and mobile sources, are considered the gold standard for measuring PM$_{2.5}$ concentrations. However, due to spatial and temporal sparsity in these monitor observations, models which estimate continuous surfaces of PM$_{2.5}$ that can be easily integrated into epidemiologic analyses are commonly utilized. A variety of modeling approaches are used for exposure estimation, including geostatistical interpolation \cite{lassman_2017} and machine learning algorithms, which can incorporate diverse information sources including remotely-sensed data products, chemical transport model output, land cover/usage variables characterizing human activity, and measurements from lower-cost and -accuracy monitors/sensors \cite{Reid_data, bi_LCS-ML_2020, Di_paper, reid_ML_2015}. Numerous model-based high-resolution PM$_{2.5}$ products exist and are used widely in air pollution (including wildfire smoke) epidemiology \cite{chen_2021, reid_maestas_2019, di_2017}.

Wildfire smoke epidemiologic studies have shown that the choice of model-based PM$_{2.5}$ estimates employed for exposure assessment can have a large impact on epidemiologic findings\cite{jiang_2023}. For instance, a study of the 2012 Washington wildfires \cite{gan_comparison_2017} found that the association between PM$_{2.5}$ and chronic obstructive pulmonary disease differed by how smoke exposure was estimated, varying both by direction (increased vs. decreased risk) and by whether or not the result was statistically significant. A study of the 2017 California wildfires \cite{cleland_2021} also found that failing to account for the uncertainty in the PM$_{2.5}$ exposure estimates resulted in an underestimation of the uncertainty in the association with health outcomes. However, to our knowledge, no studies have yet evaluated nor compared the accuracy of high-coverage (and high-resolution) PM$_{2.5}$ products for epidemiologic exposure assessment during wildfire smoke episodes. More broadly, while there has been an influx of machine learning-derived air pollution exposure datasets in recent years, most studies do not quantify uncertainty in these estimates beyond cross-validation on their training sets. Assessing whether these models perform well under a variety of scenarios is a critical first step in developing strategies to improve exposure assessment and health analyses. Our study aims to help fill this gap in the literature, as well as to explore how and why health scientists might choose exposure estimates from one model over another for a particular purpose. 

\subsection{Overview of PM$_{2.5}$ Validation and Comparison Methods}

A 2019 review evaluated several publicly-available PM$_{2.5}$ exposure products (overall, not specifically targeting wildfire smoke) and observed differences between estimates, likely due to the different input data and modeling techniques used to generate each product as well as the different spatio-temporal resolutions \cite{diao_methods_2019}. 
Of relevance to our study, the review noted that PM$_{2.5}$ in rural areas may often be overestimated, due to the fact that FRM/FEM monitors deployed by the Environmental Protection Agency (henceforth, "EPA monitors") are mostly located in urban areas, which often have more PM$_{2.5}$ from industrial and traffic emissions. By contrast, during wildfires, the concentrations of PM$_{2.5}$ are often higher in rural areas, necessitating supplemental monitoring in those areas to capture high PM$_{2.5}$ levels \cite{diao_methods_2019}.
In general, the review advocated that (a) more inter-comparison and validation of PM$_{2.5}$ datasets are needed to inform health research and public advocacy, (b) inter-comparison studies should attempt to identify methodological factors contributing to the comparison results, and (c) anticipated use(s) of the exposure data by health researchers and advocates should guide model evaluations. Our study addresses this call. 

In the context of both creating and validating wildfire-specific PM$_{2.5}$ estimates, several studies have demonstrated the importance of incorporating PM$_{2.5}$ measurements from networks of temporary monitors deployed explicitly to measure wildfire smoke exposures (more details are in the Methods section). A 2020 study estimating smoke concentrations during the October 2017 wildfires in California \cite{cleland_2020} observed that including PM$_{2.5}$ observations from these temporary monitors in addition to observations from EPA monitors in model training increased the $R^2$ by 36\%. The temporary monitoring data can also be used as independent validation data, as was done in a 2021 investigation of the 2017 Northern California wildfires \cite{oneill_multi-analysis_2021}. Although their best model compared to observations from permanent monitor locations (used in the modeling) had $R^2$ = 0.90, the highest $R^2$ compared to the temporary monitoring data (independent validation set) was 0.25. In addition to validating the estimates from one model, having an independent source of validation data is critical for comparing the accuracies of different sets of model-based estimates.

In this study, we evaluate and compare the accuracy of two publicly-available, high-coverage and high-resolution sets of daily PM$_{2.5}$ estimates spanning the western US, with an evaluation emphasis on accuracy during wildfire smoke episodes. We utilize independent validation data from temporary mobile ground monitors (deployed by the US Forest Service to capture wildfire smoke exposures) to assess accuracy of the modeled estimates across both time and space in the western US. Our findings may influence researchers' selection of a PM$_{2.5}$ exposure dataset to study health impacts of wildfire smoke and inform the creation of future exposure datasets that aim to target and/or effectively estimate PM$_{2.5}$ concentrations during wildfire events. 

\section{Materials and Methods}

\subsection{PM$_{2.5}$ Estimates Being Evaluated and Compared}

\begin{figure}[!hb]
\begin{centering}
\includegraphics[width=1\linewidth]{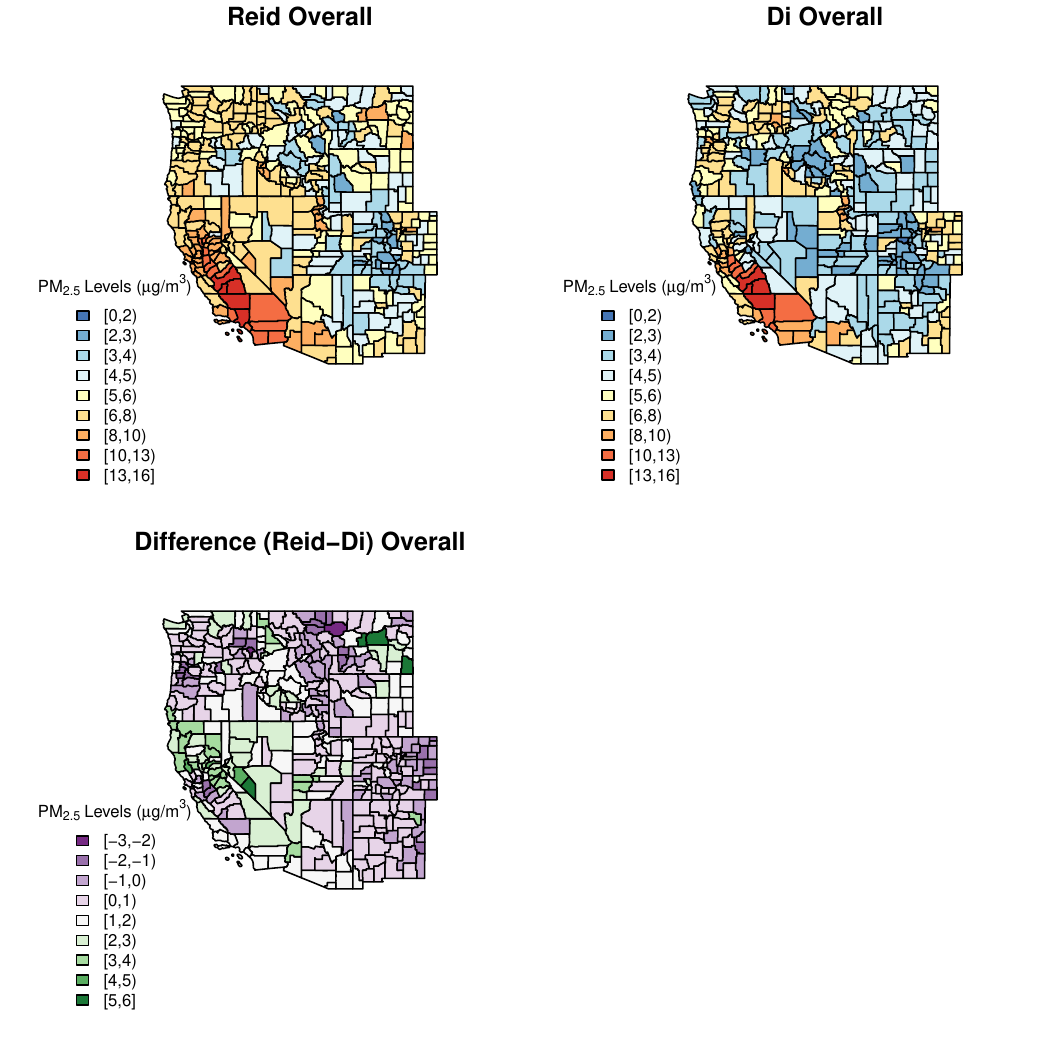} 
  \caption{Maps of county-level averages (2008-2016) for both sets of model-based estimates over the western US. The difference Reid - Di is also shown.}
  \label{fig:maps}
\end{centering}
\end{figure}

For this analysis, we compared two existing publicly-available, high-resolution and high-coverage daily PM$_{2.5}$ products, which were both created using machine learning models and which we anticipate researchers using to study the health impacts of wildfire smoke, or of air pollution more generally in the western US. County-level averages of both products, as well as the differences between them, are shown in Figure \ref{fig:maps}. Differences in seasonal averages are in the Supporting Information (SI), Figure S1. Both Figures \ref{fig:maps} and S1 exhibit clear regional and seasonal differences between the estimates from the two models.

The first set of PM$_{2.5}$ estimates evaluated here was created by Di et al. (2021); henceforth, the Di estimates \cite{Di_data}. These daily estimates are available at every 1km x 1km grid across the continental US from 2000-2016, although their 2019 paper only covers accuracy metrics through 2015. Di et al. used a geographically-weighted GAM (generalized additive model) ensemble of random forest, gradient boosting, and neural network algorithms. (Details of the PM$_{2.5}$ data and covariates used to train this model can be found in the 2019 paper\cite{Di_paper}.) This approach yielded a 10-fold cross-validation $R^2$ of 0.77 for the mountain west, 0.80 for the states on the Pacific coast, and 0.86 for the US overall. The authors noted that the model "demonstrated good performance up to 60 $\mu g / m^3$" \cite{Di_paper}. They also capped their PM$_{2.5}$ estimated concentrations at 200 $\mu g / m^3$ due to prediction instability for higher concentrations. We note that although the Di dataset was not explicitly designed for wildfire PM$_{2.5}$ exposure assessment, these estimates have been widely used in national air pollution epidemiology studies, and as previously stated, smoke from wildfires is a large source of PM$_{2.5}$ in the western US, making estimation of wildfire smoke concentrations an important part of general air pollution exposure assessment. 

The second dataset was created by Reid et al. (2021); henceforth, the Reid estimates \cite{Reid_data}. These daily PM$_{2.5}$ concentration estimates, which were created with wildfire smoke epidemiology in mind as a possible use case, are available from 2008-2018 at the centroid of every Census tract, ZIP code, and county in the western US (Arizona, California, Colorado, Idaho, Montana, Nevada, New Mexico, Oregon, Utah, Washington, and Wyoming). Reid et al. used a generalized linear model ensemble of random forest and gradient boosting tree algorithms, and achieved a 10-fold cross-validation $R^2$ of 0.73 for a 2008-2016 model and 0.72 for a 2008-2018 model. The reason for having two separate models is that one key variable, input from a chemical transport model called CMAQ (also used in the Di model), was not available past 2016 at the time of model fitting. The 2008-2018 Reid model did not use CMAQ. Unless otherwise specified, the 2008-2016 Reid estimates analyzed in this paper are from the with-CMAQ model. % clear?

One probable reason that the Di $R^2$ values are higher than those of Reid is that the monitoring dataset they used had less extreme PM$_{2.5}$ observations. Due to increasing wildfires in the western US,
more recent years have had more extreme PM$_{2.5}$ concentrations due to wildfire smoke; this increased variability is harder to accurately predict with statistical models \cite{qi_2020}. 
Indeed, both Di et al. and Reid et al. observed decreasing $R^2$ values over time when they calculated this metric by year. In addition to the Reid models covering later (more wildfire-heavy) years, they also included a wider range of monitoring observations, with some deployed near wildfires for the purpose of capturing smoke exposures.

\subsection{Validation Data}\label{ss:valdata}

We evaluated the Di and Reid model estimates using measurements from mobile monitors deployed by the AirFire research team, which is part of the US Forest Service (USFS). The AirFire monitoring data are in the Airsys and Western Regional Climate Center (WRCC) archives \cite{airsis, wrcc}, which can be obtained via the R package PWFSLSmoke \cite{smoke_package}. 

When USFS air resource advisors are assigned to wildfires in the western US (which typically occurs for large and long-duration fires and fires on federal lands), they deploy mobile monitors when and where there is concern about smoke affecting people \cite{PL_interview, EPA_report}. Specifically, the locations of the AirFire mobile monitors are normally downwind of the fires, close to where people live. Placement is at the judgement of the air resource advisor. The monitors are removed once the fire has been contained and smoke concerns have subsided (not necessarily immediately, so there are some observations in the AirFire dataset with relatively low levels of PM$_{2.5}$).

Some of these data were used to train the Reid model. Only a fraction was used because, as of data retrieval for that project (in February 2018), not all of the data was discoverable / accessible from their website, and the PWFSLSmoke package was nascent. Thus, for the current study's validation set, we removed all overlapping observations with the Reid training data (based on latitude, longitude, and date), which excluded some observations from the Airsys dataset in addition to WRCC. A map showing the locations of the monitors in the final validation set (and also indicating the number of days each was collecting data) is in the SI (Figure S2).

\subsection{Data Processing}

To compare the Di estimates and Reid estimates to the smoke monitor validation data, we identified the nearest prediction point for each dataset to each smoke monitor location based on latitude and longitude. For the Reid data, this was the nearest Census tract centroid; for the Di data, this was the nearest 1 km x 1 km grid point. We also aggregated the smoke monitor data into 24-hour averages. For any cases where there were multiple Reid estimates for the same location-day (an artifact from their data merging/imputation procedure), we took the mean of these estimates. The main validation/comparison dataset used in this paper spans 2008-2016, but we also compared the Reid data in 2017-2018 (created with the no-CMAQ model) to the smoke monitor data from those years for additional insight, as there were some very large wildfires during those two years.

To refine the smoke monitor dataset, we  removed smoke monitor observations (24-hour averages) above 1,000 $\mu g/m^3$ (n = 37, or 0.1\%) to avoid values that could be due to sensor errors. We then removed all smoke monitor observations that overlapped with the training data used by Reid et al., both (a) identical observations (n = 3,377, or 7.6\%, between 2008-2018) and (b) observations within 50 meters and within one week of an observation included in the Reid training set (n = 11,886, or 26.7\%, between 2008-2018). 

In the main 2008-2016 dataset, there were 35 location-days each from Reid and Di that were missing PM$_{2.5}$ estimates. We removed these from the analysis due to differential missingness: the smoke monitor observations on location-days missing from the Di dataset were substantially higher than those missing from the Reid dataset (the means of the smoke validation data associated with Di’s and Reid’s missing observations were
34$\mu g/m^3$ and 6$\mu g/m^3$ respectively). 

Finally, we extracted population density at each of the smoke monitor locations (by matching on Census tract) from the 2010-2014 American Community Survey using the R packages tidycensus\cite{tidycensus} and tigris\cite{tigris}. This whole procedure left 19,850 location-days across the western US (with observations available in all western states except Nevada) between 2008-2016.  
 
\subsection{Analysis}

Our validation set (from the mobile smoke monitors) contained many low PM$_{2.5}$ values in addition to very high values, often due to the monitors being left out for a while after fires were contained and smoke subsided. Thus, in this analysis we computed accuracy metrics not only for the overall validation dataset but also for particular concentration subsets of the data where the PM$_{2.5}$ concentrations exceeded certain values, corresponding to thresholds of the US Air Quality Index (AQI). The AQI has six levels ("Good", "Moderate", "Unhealthy for Sensitive Groups", "Unhealthy", "Very Unhealthy", and "Hazardous") \cite{EPA_AQI}. The PM$_{2.5}$ upper bounds for these AQI classifications are 12.1, 35.5, 55.5, 150.5, and 250.5 $\mu g / m^3$ respectively \cite{AQI_levels}. In this study, we computed accuracy metrics for the subset of location-days with smoke monitor PM$_{2.5} \geq 12.1 \mu g/m^3$ (referred to as Medium) and the subset with smoke monitor PM$_{2.5} \geq 35.5 \mu g/m^3$ (referred to as High). 
Isolating accuracy within these subsets facilitates our discussion about exposure to wildfire smoke, as we assume that (at least) the High level of PM$_{2.5}$ measured by the smoke monitors was characterizing wildfire smoke days, and the Medium level might also.

In addition to summary statistics for each of these subsets and for each state in the western US, we compared the Di and Reid estimates to the smoke monitor observations using the following metrics: mean bias, normalized mean bias, median absolute difference, median ratio, mean ratio, root-mean squared difference (RMSD), squared correlation ($R^2$), spatial RMSD, temporal RMSD, spatial correlation, and temporal correlation. Formulas for all these metrics, which have been used in similar studies \cite{ye_2021, oneill_multi-analysis_2021, jin_comparison_2019}, are in the SI. Briefly, the spatial metrics are calculated after averaging across all days at each location, and they inform about the accuracy with which each set of estimates captures the spatial patterns in wildfire-related PM$_{2.5}$. The temporal metrics are calculated after averaging across all locations on each day, and they inform about the accuracy with which temporal patterns of wildfire-related PM$_{2.5}$ are captured by the estimates. 

Because environmental health studies are concerned with population exposure, the main analyses in this paper are weighted by the population density of the census tracts in which the smoke monitors were located. Unless otherwise stated, we henceforth refer to population density-weighted metrics. As sensitivity analyses we include metrics unweighted by population density in the SI. 

To investigate the possibility of differential model performance in the warm season (when the majority of wildfires occur in the western US), we calculated the pairwise mean bias, median ratio, RMSD, $R^2$, and both spatial and temporal correlation between the Reid estimates, Di estimates, and smoke monitor observations, stratified by whether the observation was from the warm season or not, which we defined as the months May through October. 

To gain additional insights, we calculated the main metrics for the Reid estimates compared with the smoke monitor observations from 2017-2018 -- years for which there are no Di estimates (n = 9,662). Lastly, we calculated these metrics for both the Di estimates and Reid estimates compared with the EPA 
monitor observations (2008-2016), on which both models were trained (n = 927,414). This allows for comparison of how well the models perform in general over the western US (not just focusing on wildfires).

Finally, we investigated the accuracy of both models' classifications of PM$_{2.5}$ according to the AQI. This was motivated by the possibility that health researchers might be inclined to use categorical smoke exposures to address the problem of large-magnitude model-based estimation errors at high concentrations of PM$_{2.5}$. Indeed, use of both binary \cite{liu_2017, reid_2016} and multi-category \cite{jones_2020, alman_2016} exposures, including AQI classifications \cite{hutchinson_2018}, are not unusual in the wildfire smoke epidemiology literature. Another motivation for our AQI-based analysis is that the AQI is now often used to trigger interventions to protect the public from wildfire smoke \cite{holm_2021}. As part of this analysis, we compared the AQI classifications of the Di and Reid model estimates to the AQI classifications of the smoke monitor observations. We also calculated the overall rates of misclassification, under- and over-classification (e.g. the observed AQI was "Moderate" and the model predicted "Unhealthy for Sensitive Groups"), and misclassification by more than one level of the AQI (e.g. the observed AQI was "Unhealthy for Sensitive Groups" and the model estimated "Good"). 

In addition to using all six AQI classes, we also investigated using a binary exposure (non-smoke = "Good" or "Moderate" vs. smoke = "Unhealthy for Sensitive Groups" or above). We calculated a confusion matrix as well as sensitivity, specificity, positive predictive value, and negative predictive value for each dataset compared to the smoke monitor observations.

\section{Results}

\subsection{Summary Statistics}

\begin{table}[!h]
\centering
\begin{tabular}{|p{2cm} | ccc | ccc | ccc |}
  \hline
   & \multicolumn{3}{c|}{\textbf{Overall (N = 19,850)}}  & \multicolumn{3}{c|}{\textbf{Medium (N = 5,432)}} & \multicolumn{3}{c|}{\textbf{High (N = 1,034)}} \\
  \hline
\textbf{Metric} & Monitor & Di & Reid & Monitor & Di & Reid & Monitor & Di & Reid \\ 
  \hline
 Minimum & 0.00 & 0.00 & 0.00 & 12.12 & 0.00 & 0.63 & 35.50 & 0.00 & 0.63 \\ 
   \hline
1st \mbox{Quartile} & 4.38 & 3.92 & 6.36 & 14.58 & 6.81 & 10.89 & 45.38 & 6.49 & 18.41 \\ 
   \hline
Median & 7.58 & 6.34 & 8.51 & 18.54 & 11.43 & 14.68 & 65.46 & 10.92 & 27.67 \\ 
   \hline
Mean & 12.11 & 7.92 & 10.82 & 29.50 & 12.44 & 17.84 & 95.10 & 17.33 & 33.86 \\ 
   \hline
3rd \mbox{Quartile} & 12.42 & 10.28 & 12.49 & 26.04 & 15.34 & 19.14 & 69.92 & 23.16 & 36.89 \\ 
   \hline
Maximum & 928.04 & 190.71 & 227.11 & 928.04 & 190.71 & 227.11 & 928.04 & 190.71 & 227.11 \\ 
   \hline
Standard \mbox{Deviation} & 30.03 & 6.42 & 9.79 & 55.34 & 9.16 & 15.59 & 129.44 & 17.97 & 32.14 \\ 
   \hline
\end{tabular}
\caption{\label{tab: main-summary} Summary statistics (all in $\mu g/m^3$) of the smoke monitor observations, Di estimates, and Reid estimates in the merged validation set, subset by level of PM$_{2.5}$ and weighted by population density.}
\end{table}

Table \ref{tab: main-summary} shows how the smoke monitor observations not only tend to have higher PM$_{2.5}$ values than either of the model estimates, but also have higher standard deviations, in each of the three levels of PM$_{2.5}$. These patterns become more pronounced as the level of PM$_{2.5}$ increases. Between the two models, the Reid estimates have higher summary statistics than the Di estimates, putting their values closer to those of the smoke monitor summary statistics (the exception to the latter is first quartile overall PM$_{2.5}$).

When we consider the mean and standard deviation stratified by state of the smoke monitor observations and model estimates (Table S1), similarly to Table \ref{tab: main-summary}, the mean smoke monitor observations tend to be higher than the mean estimates from either model (universally for medium and high levels of PM$_{2.5}$), and the standard deviations tend to be higher as well. For the most part, the mean Reid estimates are substantially higher than the mean Di estimates. The only exceptions to this where the Di mean is more than 1$\mu g/m^3$ higher than the Reid mean are Montana's high PM$_{2.5}$ (Di = 48$\mu g/m^3$  vs Reid = 41$\mu g/m^3$) and overall PM$_{2.5}$ (Di = 10$\mu g/m^3$ vs Reid = 8$\mu g/m^3$) and Wyoming's high PM$_{2.5}$ (Di = 29$\mu g/m^3$  vs Reid = 26$\mu g/m^3$).  

% A final observation is that the discrepancy between the smoke monitor and model means tends to be lower for states with more smoke monitor observations. This could simply be due to more relatively-low concentrations (which are easier for the models to predict) being measured in those states. (Note that the ranking of the states based on the amount of data they had to train on is not necessarily the same as the ranking based on the amount of data for each state in the smoke monitor dataset.)

\subsection{Scatterplots by Year and Season}

\begin{figure}[!hb]
\begin{centering}
\includegraphics[width=1\linewidth]{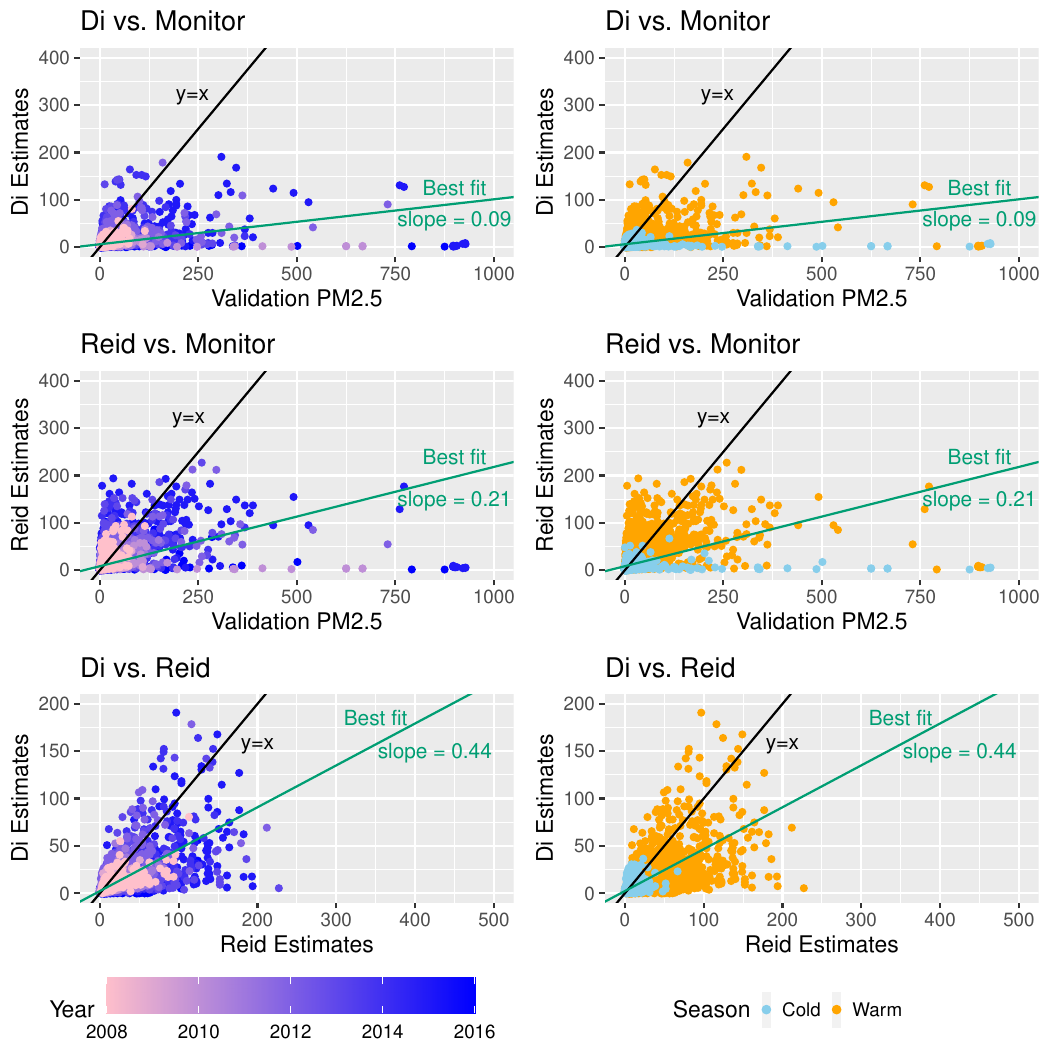} 
  \caption{Pairwise comparisons of the Di estimates, Reid estimates, and the smoke monitor observations (all in $\mu g/m^3$), with points colored by year and by season. (We define Warm Season as May through October.) Slopes of the best-fitting Deming lines are shown for each.}
  \label{fig:scatter}
\end{centering}
\end{figure}

Figure \ref{fig:scatter} shows pairwise comparisons of the Di, Reid, and smoke monitor data, with points colored by year in the left column and by season in the right column. We first note that both the Di and Reid model estimates have a large number of under-predictions at high values of smoke monitor PM$_{2.5}$. The Di estimates are almost entirely under-predictions for smoke monitor values $>125\mu g/m^3$ and the Reid estimates are almost entirely under-predictions for smoke monitor values $>200\mu g/m^3$. These under-predictions contribute to the Deming lines of best fit (shown in green) having flatter slopes than the y=x lines (shown in black). Comparing the slopes of the green and black lines, the larger slope on the Reid vs. Monitor plots (0.21) compared to the Di vs. Monitor plots (0.09) indicates that the Reid estimates are better capturing the variation in the smoke monitor observations, particularly at higher concentrations. However, we also see from the Di vs. Reid plots (slope of best fit line = 0.44) that the two models' estimates are closer to each other than either of them is to the smoke monitor observations.

The scatterplots colored by year of observation illustrate how more extreme smoke monitor observations have been collected in more recent years. The scatterplots colored by season (May-October compared to November-April) offer a couple of insights. Unsurprisingly, the majority of observations in the smoke monitor dataset are from the warm months. Also, the observations from the cold months tend to be clustered closer to the origin, except for a string of high concentrations that both models appear to struggle to capture -- though the Reid model identifies slightly more of this variability.

\subsection{Comparison Metrics Results}

\begin{table}[!h]
\centering
\begin{tabular}{|p{3.6cm} | p{1.5cm} p{1.5cm} |  p{1.5cm} p{1.5cm} | p{1.5cm} p{1.5cm}|}
  \hline
  & \multicolumn{2}{p{3cm}|}{\textbf{Overall \mbox{(N = 19,850)}}}  & \multicolumn{2}{p{3cm}|}{\textbf{Medium \mbox{(N = 5,432)}}} & \multicolumn{2}{p{3cm}|}{\textbf{High \mbox{(N = 1,034)}}} \\
  \hline
\textbf{Metric \mbox{(Optimal Value)}} & Di vs. \mbox{Monitor} & Reid vs. \mbox{Monitor} & Di vs. \mbox{Monitor} & Reid vs. \mbox{Monitor} & Di vs. \mbox{Monitor} & Reid vs. \mbox{Monitor} \\ 
  \hline
 Mean Bias (0) & -4.19 & -1.29 & -17.06 & -11.66 & -77.77 & -61.24 \\ 
   \hline
Normalized \mbox{Mean Bias (0)} & -0.35 & -0.11 & -0.58 & -0.40 & -0.82 & -0.64 \\ 
   \hline
\mbox{Median Absolute} \mbox{Difference} (0) & 3.01 & 2.66 & 8.44 & 6.17 & 48.48 & 35.03 \\ 
   \hline
\mbox{Median Ratio (1)} & 0.85 & 1.15 & 0.55 & 0.73 & 0.14 & 0.46 \\ 
   \hline
 Mean Ratio (1) & 7.63 & 13.38 & 0.60 & 0.79 & 0.28 & 0.50 \\ 
   \hline
RMSD (0) & 29.28 & 27.83 & 57.18 & 54.07 & 150.31 & 142.44 \\ 
   \hline
$R^2$ (1) & 0.07 & 0.15 & 0.02 & 0.09 & 0.00 & 0.01 \\ 
   \hline
\mbox{Spatial RMSD (0)} & 74.74 & 71.49 & 85.52 & 81.37 & 154.77 & 148.82 \\ 
   \hline
\mbox{Temporal RMSD (0)} & 19.18 & 18.51 & 69.61 & 68.42 & 153.94 & 148.75 \\ 
   \hline
Spatial \mbox{Correlation (1)} & 0.35 & 0.39 & 0.23 & 0.33 & 0.00 & 0.15 \\ 
   \hline
Temporal \mbox{Correlation (1)} & 0.27 & 0.34 & 0.04 & 0.10 & -0.09 & -0.04 \\ 
   \hline
\end{tabular}
\caption{\label{tab: main-metrics} Metrics (weighted by population density) comparing the Di estimates and Reid estimates to the smoke monitor observations (2008-2016), stratified by PM$_{2.5}$ level. All metrics except for median ratio, mean ratio, $R^2$, and spatial/temporal correlation are in $\mu g/m^3$.}
\end{table}

The mean bias, normalized mean bias, and median ratio values comparing the Di and Reid estimates to the smoke monitor observations (in Table \ref{tab: main-metrics}) confirm that both models tend to underestimate PM$_{2.5}$ during smoke events. For instance, for high PM$_{2.5}$, the mean bias for the Di estimates is -77.8$\mu g/m^3$ and for the Reid estimates is -61.2$\mu g/m^3$. The glaring contradiction is the mean ratio for overall PM$_{2.5}$, which is substantially greater than 1 for both models (7.6 for Di and 13.4 for Reid). This is heavily influenced by instances when the monitor observation is near-zero and the models predict medium-low concentrations, which yields very high ratios. The Reid metric in particular is affected by 18 instances where the monitor observations are in their lowest quartile ($ < 4\mu g/m^3$) and the model estimates are high ($ > 35.5 \mu g/m^3$). More details on these large over-predictions are in the SI. By contrast, the Di estimates do not exhibit this issue with large over-predictions when the smoke monitor observations are near-zero.

Across all metrics, both models' performance deteriorates (even in normalized terms) as PM$_{2.5}$ increases. The large RMSD values and small $R^2$ values indicate that both models struggle to capture high values of PM$_{2.5}$, which in this validation set are primarily due to wildfire smoke. Indeed, at high PM$_{2.5}$ concentrations, the estimates from both models are effectively uncorrelated with the smoke monitor observations. Compared to the smoke monitor observations, the only metric in Table \ref{tab: main-metrics} for which the Di estimates clearly outperform the Reid estimates is the overall mean ratio (which, as mentioned above, is heavily impacted by discrepancies at lower levels of PM$_{2.5}$ as measured by the smoke monitors). 

A result that is harder to interpret is that for high PM$_{2.5}$, both models' temporal correlations are negative (-0.09 for the Di estimates and -0.04 for the Reid estimates), which indicates that higher monitor observations are linearly associated with lower model estimates. Negative temporal correlation could be due to the time it takes for wildfire smoke to travel away from a fire area and be measured by EPA monitors in other areas, the data from which constituted the majority of the training sets for the Di and Reid models. However, these temporal correlation values have relatively small magnitudes, so we caution against over-interpreting these results.  

\begin{figure}[!h]
\begin{centering}

\begin{tabular}{|cc|cc|cc|}
  \hline
Overall Warm & Overall Cold & Medium Warm & Medium Cold & High Warm & High Cold\\ 
\hline
  15,570 &  4,280 &  4,539 &   893 &   945 &    89 \\ 
   \hline
\end{tabular}

\includegraphics[width=1\linewidth]{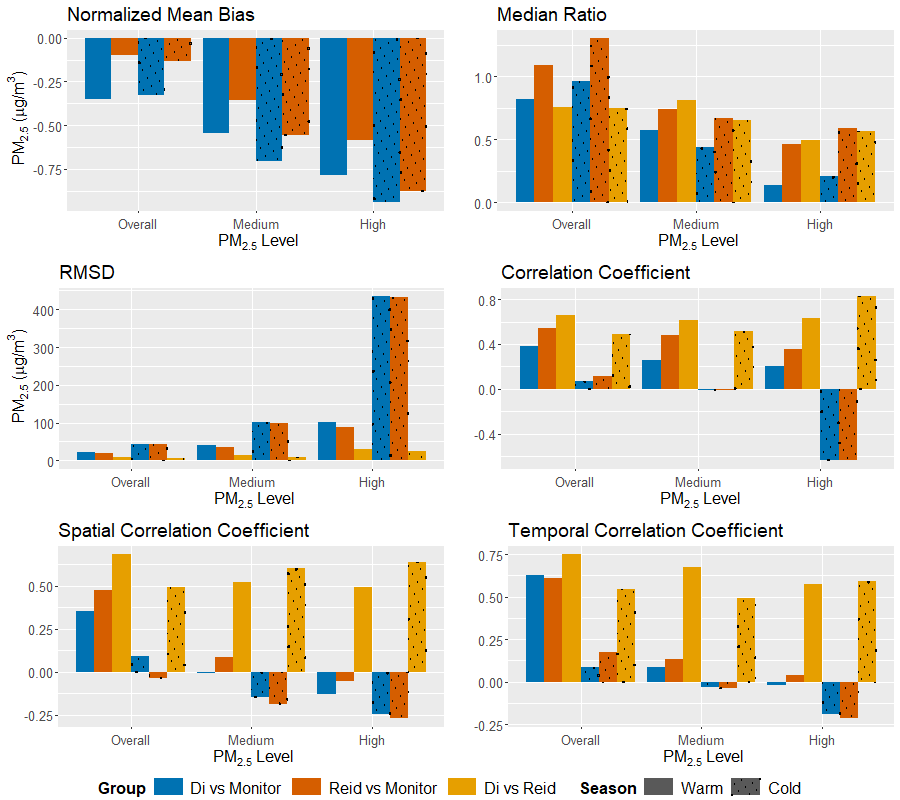} 

\caption{Pairwise normalized mean bias, median ratio, RMSD, $R^2$, and spatial and temporal correlations for the smoke monitor data, Di estimates and Reid estimates, stratified by PM$_{2.5}$ level and whether the month was between May and October (which we define as the warm season in the western US) and weighted by population density. The plot for normalized mean bias does not include metrics for Di vs Reid to avoid having to use one of the sets of model estimates for normalization. Sample sizes for each subset are shown at the top.}
  \label{fig:WS-plots}
\end{centering}
\end{figure}

Figure \ref{fig:WS-plots} shows pairwise normalized mean bias, median ratio, RMSD, correlation, and spatial and temporal correlations for the Di and Reid estimates and smoke monitor observations, stratified by level of monitor PM$_{2.5}$ and warm/cold season. Note that better agreement for each pair is indicated by normalized mean bias and RMSD values closer to zero, and median ratio, and (regular, spatial and temporal) correlation values closer to one. Clearly, there are more discrepancies between the three datasets as the level of PM$_{2.5}$ increases and as we move from warm to cold season. The generally worse accuracy of the model estimates in the cold season may have implications for studies investigating differential health impacts of wild and prescribed fires, because the latter mostly occur/have been increasingly occurring during what we are defining as the cold season \cite{baijnath-rodino_2022, ryan_2013}. Worse estimation accuracy for smoke during the cold season may also have implications for studies investigating health interaction effects between temperature and smoke (or PM$_{2.5}$ in general), for which there is already some evidence \cite{reid_2016, anenberg_2020}.

The yellow bars, comparing the Di and Reid estimates, show by far the highest agreement (across levels of PM$_{2.5}$ and warm/cold season), confirming that the two sets of model estimates tend to be more similar to each other than either is to the smoke monitor data. However, median ratio for overall PM$_{2.5}$ indicates closer agreement between model estimates and smoke monitor observations than between the two models.

With the exception of temporal correlation for overall PM$_{2.5}$ in the warm season, the only subset-metrics for which the Di estimates outperform the Reid estimates (compared with the smoke monitor observations) occur in the cold season.  Di's median ratio for overall PM$_{2.5}$ is closer to one and both the temporal correlation for high PM$_{2.5}$ and the spatial correlations across all levels of PM$_{2.5}$ are less negative than their Reid counterparts. As noted previously, negative temporal correlation could be due to the time it takes for wildfire smoke to travel away from a fire area and be measured by EPA monitors in other areas. Spatially, negative correlation could be due to the aforementioned issue with the majority of EPA monitors being placed in more urban areas with higher background PM$_{2.5}$ while the majority of wildfires and smoke monitors are located in more rural areas. While this is an interesting result, keep in mind that our sample size for high PM$_{2.5}$ in the cold season is relatively small (n = 89). In any case, it appears that the Reid estimates struggle to capture spatio-temporal variability of wildfire smoke exposures more in the cold season but generally are able capture more variability than the Di estimates in the warm season, across levels of PM$_{2.5}$.

We present and discuss the results of the AQI classification and binary smoke / nonsmoke classification analyses in the Supplemental Notes (in the SI). The results of the sensitivity analyses (metrics unweighted by population density, comparing the Reid estimates in 2017-2018 with the smoke monitor validation data, and comparing the Reid and Di estimates with the EPA monitor observations from 2008-2016) are also in the Supplemental Notes.

\section{Discussion}

In this project, we evaluated and compared the modeled daily PM$_{2.5}$ concentrations over the western US in 2008-2016 from two publicly-available high-resolution datasets, with an eye towards their suitability for studying the health impacts of wildfire smoke. 

Using independent validation data from mobile monitors deployed for the specific purpose of measuring wildfire smoke concentrations near where people live, we found that across levels of PM$_{2.5}$, the smoke monitor validation observations were higher than the estimates from either model. This issue gets worse as the level of PM$_{2.5}$ increases (even on a normalized scale). The relatively poor performance of both models compared to the smoke validation data (overall, Di RMSD = 29.3 $\mu g / m^3$ and $R^2$ = 0.07; Reid RMSD = 27.8 $\mu g / m^3$ and $R^2$ = 0.15) indicates that relying on estimates from machine learning models of air pollution concentrations to analyze the health impacts of wildfire smoke could introduce substantial exposure measurement error. Of course, only using monitoring data would likely result either in a much smaller sample size for health analyses (if one only considered people living in immediate vicinity of a monitor) or more exposure measurement error than the modeled estimates \cite{keller_2019, ebelt_2022}, especially for areas without a monitor close by.

Our analysis also highlights the importance of using independent smoke-specific ground observations for evaluating model estimates: the cross-validation $R^2$ values reported in the Di and Reid papers were between four and eleven times better than the $R^2$ values compared to the smoke monitor observations. 
This discrepancy aligns with others' findings \cite{oneill_multi-analysis_2021}.

In general, our comparison metrics show that the Reid estimates outperform the Di estimates when estimating PM$_{2.5}$ as measured by the smoke monitors, and this is true across levels of PM$_{2.5}$. However, the Di estimates perform slightly better than the Reid estimates at capturing spatio-temporal variability during the cold season, when more prescribed fires occur. That said, both models perform much worse during the cold season than during the warm season, likely in part because there are far fewer wildfires (and thus we have far less smoke monitor data) from those months.

For context, when compared with observations from the EPA monitors (on which both models were trained), the Reid estimates capture more large-scale variability in western US PM$_{2.5}$ than the Di estimates. However, a few metrics suggest that the Di model is capturing a bit more localized variability, likely from non-wildfire (i.e. urban / industrial) sources of PM$_{2.5}$, while not performing as well on the large-scale variability, which is often influenced by wildfire smoke.

We conclude that if researchers were to use one of these datasets for investigating wildfire smoke exposure and health impacts over the western US, the Reid estimates would be preferable to the Di estimates. However, both sets of estimates have their drawbacks. Any researchers interested in using estimates from these datasets (or similar datasets) must address the challenge of modeled data substantially underestimating peak PM$_{2.5}$ events. Without adjustment, it is likely that under-estimation of these high PM$_{2.5}$ concentrations would bias health effects towards the null \cite{wei_2022}.

An approach that we imagine health researchers might consider to address the issue of large-magnitude errors is aggregating PM$_{2.5}$ estimates into categorical exposures. To address this possibility, we investigated classification accuracies under both the six-level AQI and a binary smoke/non-smoke classification. Overall, the Reid and Di models misclassify the AQI (as measured by the smoke monitors) 20.7\% and 23.2\% of the time, respectively. However, the Di estimates exhibit more under-classification of the AQI, whereas the Reid estimates exhibit more over-classification. It is unclear whether these two cases (over- and under-classification) might differentially bias health effect estimates based on a categorical PM$_{2.5}$ exposure. For the binary exposure, both models misclassify only 3.4\% of the time, and are able to identify non-smoke days over 95\% of the time. However, the Reid model is able to identify a higher fraction of the smoke days (sensitivity) while if the Di model classifies a smoke day, then there is a higher probability that it was smoky (positive predictive value).  

One of the likely reasons that the Reid estimates are able to capture higher wildfire smoke concentrations is the inclusion of additional wildfire-specific monitoring observations in their training data. Our usage of the mobile smoke monitors for independent validation of both models further illustrates the value of such data. As presented in the Supplemental Notes, the Reid estimates from 2017-2018 perform much better than those from 2008-2016 (compared to the smoke monitor observations). We infer that this is because there has been increased collection of these wildfire-specific data over time. Before we processed the data (e.g., to remove overlap with the Reid training set), there was an approximately linear increase in the mobile smoke monitor measurements between 2008 and 2018, with an average increase of 736 daily observations per year. 
We recommend broader usage of such data in future wildfire smoke exposure modeling and validation analyses. In addition to helping estimate high concentrations, this may help to address the issue of negative temporal correlation (which both the Di and Reid estimates exhibited for high levels of PM$_{2.5}$ observed by the smoke monitors).

Our conclusion that more western wildfire-specific monitoring is needed to capture population exposure to wildfire smoke is in agreement with the findings of two studies published in 2022, which used very different methodologies -- one used mode decomposition on the Di estimates\cite{kelp_2022} and the other used chemical transport modeling under past and future climate conditions\cite{marlier_2022}. Both of these studies focused on the network of EPA monitors (and our study additionally used the USFS mobile smoke monitors), but another possibility for future years is including data from low-cost air quality sensors, which are much less accurate than EPA monitors and the mobile smoke monitors, but can provide more high-resolution spatial and temporal information on PM$_{2.5}$ \cite{bi_LCS-ML_2020}. During the 2021 fire season, the USFS deployed low-cost sensors to fill a shortage in the smoke monitors \cite{PL_interview}. EPA researchers have also developed a smoke correction for PurpleAir sensors \cite{PA_smoke_correction}, a common brand of low-cost PM$_{2.5}$ sensors that many people have purchased for use in their homes and communities. However, including lower-accuracy air quality observations from these low-cost sensors can introduce new challenges \cite{bi_LCS-ML_2020}.

In addition to training on wildfire smoke-specific monitor data, other aspects of the Reid model that may have helped it better capture high PM$_{2.5}$ from wildfire events are utilization of a satellite product identifying active fire points and inclusion of spatio-temporal indicator variables for state and subregions of the western US, which may have helped optimize the model for this region.
The Di model, in addition to covering the entire US, incorporated more spatial and temporal smoothing. 
Specifically, they used spatially lagged monitored PM$_{2.5}$, as well as one-day lagged and three-day lagged values of spatially lagged terms, as predictor variables in their final model.
While this technique may be beneficial for estimation of PM$_{2.5}$ overall, it might also smooth out some of the extreme, short-term spikes due to wildfires. In future, we suggest that researchers modeling the effects of wildfire smoke (or other regionally-clustered phenomena) consider adding more event- and/or region-specific information. Users of model-based air pollution estimates should also be mindful of prediction caps (e.g., the Di estimates were capped at 200 $\mu g/m^3$ whereas the Reid estimates were not capped) or other means of "taming" large estimates, which can be inappropriate in the wildfire smoke context.

A drawback of region-specific modeling is that wildfire smoke can travel large distances. As suggested by a 2021 study \cite{odell_2021}, diluted smoke concentrations in more populated areas (i.e. in the eastern US) may increase attributable mortality and asthma morbidity from wildfire smoke relative to higher smoke concentrations in less populated areas (i.e. much of the western US). While the Di estimates cover the continental US, the Reid estimates only cover the western US.

There is also a difference in spatial resolution between the two datasets: the Di estimates are available at every 1 km x 1 km grid cell, while the Reid estimates are available at the centroid of every county, ZIP code, and Census tract. Given that most health data is aggregated to one of these geographical units and many of the covariates used to train the models were coarser than 1 km x 1 km, this difference may not greatly affect health analyses. On the other hand, when researchers have address information or are linking in other gridded data, the Di estimates may be preferred. In any case, the intended application of the PM$_{2.5}$ estimates should inform their use, in epidemiologic studies or otherwise. 

Another important aspect of both the Reid and Di models is that they estimated total PM$_{2.5}$, not smoke-specific PM$_{2.5}$ (total minus background PM$_{2.5}$). To investigate exposure to and health impacts of wildfire smoke, researchers must decide how to classify "smoke days" and "non-smoke days". However, this is also necessary for groups that choose to model the counterfactual background PM$_{2.5}$ so that they can calculate the amount that is wildfire-specific. And, from a health impacts standpoint (e.g. for estimation of concentration-response functions), people are experiencing total PM$_{2.5}$ and therefore any health impacts are influenced not just by the wildfire PM$_{2.5}$ but total PM$_{2.5}$. In future, using component-specific PM$_{2.5}$ estimates may help improve characterization of wildfire smoke estimates and subsequent calculation of health effects \cite{stowell_2020}. 

A final note about both models is that although we can estimate predictive accuracy using cross-validation and comparison with independent and/or smoke-specific monitoring data, these machine learning approaches do not allow for quantifying the uncertainty of individual predictions, which might then be used in subsequent estimation of health effects (i.e., health impact analyses or risk assessments) using an existing concentration-response function. Future air pollution modelers might consider a method such as Bayesian Additive Regression Trees (BART), which in addition to being a flexible ensemble modeling approach similar to those used in many air pollution predictive models, allows for natural estimation of predictive uncertainties using Bayesian modeling techniques \cite{zhang_2020}. On the flip side, several health impact assessments (including one wildfire-specific study) have observed that the magnitude of mortality and morbidity estimates is much more dependent on the uncertainty of the concentration-response function being used than by uncertainty in the choice of PM$_{2.5}$ estimates to which the concentration-response function is applied \cite{jin_comparison_2019, cleland_2021}. This is good if one is applying a robust concentration-response function to a noisier exposure dataset, but bad if one is attempting to develop a wildfire smoke-specific concentration-response function using exposure estimates from models which struggle to predict high concentrations of PM$_{2.5}$, which will likely lead to more uncertainty in the resulting concentration-response function.

The primary limitation of our main analysis is that the smoke monitor observations were not evenly spread across our study region in either space or time, as illustrated by Figure S2. For instance, there were zero observations available from Nevada, and 2017-2018 had 50\% the number of observations that 2008-2016 (a nine-year period) had. This could imply that we are monitoring air quality more routinely or just that we are monitoring more during heavy wildfire events in the western US, which have been increasing \cite{odell_2019}. The heterogeneous distribution of the smoke monitor observations increased the difficulty of interpreting some of the metrics we used in our analysis, such as spatial and temporal correlation/RMSD. For instance, since most of the smoke monitors were collecting data during the summer months, spatial correlation in this analysis likely includes some temporal information. Another limitation of our analysis is that the warm and cold seasons (at least as they relate to wild and prescribed fires) are not exactly the same for all the states in the western US \cite{ryan_2013}, so our definitions of the warm season as May-October and the cold season as November-April may have obscured some of the seasonal heterogeneity across states. 

\subsection{Code and Data Availability}

Code used to process and analyze the data and visualize the results can be found at https://github.com/haohaojy/wildfire-PM2.5-comparison. The processed datasets used in this analysis are available on Harvard Dataverse: https://doi.org/10.7910/DVN/MBAVER. The raw datasets (prior to our processing) are also all publicly available.

\subsection{Funding}

\begin{itemize}
    \item NIH grant 5T32ES007142 (EMC)
    \item NIH grant K01ES032458 (RCN)
    \item Harvard Climate Change Solutions Fund (RCN)
\end{itemize}

\subsection{Authors' Contributions} 

\begin{itemize}
    \item Conceptualization: all authors.
    \item Data processing: JH, EMC
    \item Data analysis: EMC, JH
    \item Data visualization: EMC
    \item Drafting manuscript: EMC, JH
    \item Editing manuscript: EMC, RCN, CER, PD, DB
    
\end{itemize}

\subsection{Conflicts of Interest}

Disclaimer: CER and EMC were the primary creators of the Reid estimates. However, DB and RCN also have close ties to the research group that created the Di estimates. Otherwise, we have no conflicts of interest to declare.

\begin{acknowledgement}

The authors thank:
\begin{itemize}
    \item The National Studies on Air Pollution and Health (NSAPH) research lab for their support of this project.
    \item Pete Lahm of the Interagency Wildland Fire Air Quality Response Program for answering all of our questions about the mobile smoke monitors.
    \item The AirFire Research Team, specifically Amy Marsha, for helping us obtain the Airsys and WRCC data.
\end{itemize}

\end{acknowledgement}

\begin{suppinfo}

\begin{enumerate}
    \item Equations for the metrics used to compare the smoke monitor observations, Di estimates, and Reid estimates
    \item Supplemental Notes: 
    \begin{enumerate}
        \item Investigating Large Mean Ratios
        \item Results of the AQI Classification Analysis
        \item Results of the Binary Smoke / Nonsmoke Classification Analysis
        \item Results of the Sensitivity Analysis Unweighted By Population Density
        \item Results of the Sensitivity Analysis on the Reid Estimates without CMAQ (Including 2017-2018)
        \item Results of the Sensitivity Analysis Comparing Reid and Di to EPA Monitor Observations
    \end{enumerate}
    \item Figure S1: Maps of the seasonal differences between county-level averages of the Di and Reid estimates (2008-2016)
    \item Figure S2: Map of all monitoring observations in the validation set, colored by the number of days they were deployed
    \item Table S1: Mean (standard deviation) for each data source in the validation set, by state and level of PM$_{2.5}$
    \item Table S2: Comparison metrics (like Table 2) unweighted by population density
    \item Table S3: Frequency tables of each dataset’s classification of the AQI in the validation set
    \item Table S4: AQI classification metrics for the Di and Reid estimates compared with the smoke monitor observations
    \item Table S5: Frequency tables of each dataset’s binary classification of AQI ("Non-smoke" vs. "Smoke") in the validation set
    \item Table S6: Summary statistics for the Di and Reid estimates' binary classification of AQI ("Non-smoke" vs. "Smoke") compared with the smoke monitor observations
    \item Table S7: Comparison metrics (like Table 2) for the validation observations and the Reid estimates without CMAQ, split into 2017-2018 and 2008-2016
    \item Table S8: Comparison metrics (like Table 2) for both sets of model estimates and the EPA monitor observations
\end{enumerate}

\end{suppinfo}

\bibliography{wildfire_pm2-5}



\section*{Supplementary material}

Table S3: Frequency tables for the Di and Reid estimates' classification of the AQI in the validation set. Levels 1-6 of the AQI are described in Methods: Analysis.

|                   | Di AQI Class |      |     |     |     |      |           |
|-------------------|-------|------|-----|-----|-----|------|-----------|
| Monitor AQI Class | 1     | 2    | 3   | 4   | 5   | 6    | Marginals |
| 1                 | 13801 | 609  | 6   | 2   | 0   | 0    | 14418     |
| 2                 | 3289  | 1040 | 47  | 22  | 0   | 0    | 4398      |
| 3                 | 219   | 209  | 24  | 25  | 0   | 0    | 477       |
| 4                 | 168   | 174  | 51  | 20  | 3   | 0    | 416       |
| 5                 | 29    | 27   | 9   | 17  | 1   | 0    | 83        |
| 6                 | 31    | 9    | 3   | 13  | 2   | 0    | 58        |
| Marginals         | 17537 | 2068 | 140 | 99  | 6   | 0    | 19850     |
| Percent Change (%)| +22   | -53  | -71 | -76 | -93 | -100 |           |

(a) Di classifications compared to smoke monitor classifications

|                   | Reid AQI Class |      |     |     |     |      |           |
|-------------------|-------|------|-----|-----|-----|------|-----------|
| Monitor AQI Class | 1     | 2    | 3   | 4   | 5   | 6    | Marginals |
| 1                 | 12976 | 1321 | 85  | 36  | 0   | 0    | 14418     |
| 2                 | 2713  | 1327 | 211 | 145 | 2   | 0    | 4398      |
| 3                 | 115   | 193  | 78  | 89  | 2   | 0    | 477       |
| 4                 | 72    | 126  | 87  | 127 | 4   | 0    | 416       |
| 5                 | 16    | 6    | 9   | 43  | 9   | 0    | 83        |
| 6                 | 23    | 5    | 3   | 22  | 5   | 0    | 58        |
| Marginals         | 15915 | 2978 | 473 | 462 | 22  | 0    | 19850     |
| Percent Change (%)| +10   | -32  | -1  | +11 | -73 | -100 |           |

(b) Reid classifications compared to smoke monitor classifications

Table S4: AQI classification metrics for the Di and Reid estimates compared with the smoke monitor observations, weighted and unweighted by population density.

| Comparison               | Misclassified | Underclassified | Overclassified | Large Misclass. |
|--------------------------|---------------|-----------------|----------------|-----------------|
| Di vs Monitor, weighted  | 23.22         | 15.75           | 7.47           | 2.58            |
| Reid vs Monitor, weighted| 20.69         | 11.02           | 9.67           | 2.17            |
| Di vs Monitor, unweighted| 25.01         | 21.41           | 3.60           | 3.59            |
| Reid vs Monitor, unweighted| 26.87       | 17.32           | 9.55           | 3.36            |

Table S5: Frequency tables for the Di and Reid estimates' binary classification of AQI as Non-smoke ("Good" or "Moderate") vs. Smoke ("Unhealthy for Sensitive Groups" and above) in the validation set.

|  | Di Binary Exposure | | |
|---|---|---|---|
| Monitor Binary Exposure | Non-smoke | Smoke | Marginals |
| Non-smoke | 18739 | 77 | 18816 |
| Smoke | 866 | 168 | 1034 |
| Marginals | 19605 | 245 | 19850 |

(a) Di classifications compared to smoke monitor classifications

|  | Reid Binary Exposure | | |
|---|---|---|---|
| Monitor Binary Exposure | Non-smoke | Smoke | Marginals |
| Non-smoke | 18337 | 479 | 18816 |
| Smoke | 556 | 478 | 1034 |
| Marginals | 18893 | 957 | 19850 |

(b) Reid classifications compared to smoke monitor classifications

Table S6: Summary statistics for the Di and Reid estimates' binary classification of AQI as Non-smoke ("Good" or "Moderate") vs. Smoke ("Unhealthy for Sensitive Groups" and above) in the smoke monitor validation set, weighted and unweighted by population density. PPV = positive predictive value, NPV = negative predictive value.

| Comparison | Misclassified | Sensitivity | Specificity | PPV | NPV |
|---|---|---|---|---|---|
| Di vs Monitor, weighted | 3.44 | 10.51 | 99.81 | 67.50 | 96.73 |
| Reid vs Monitor, weighted | 3.44 | 28.24 | 99.13 | 55.19 | 97.34 |
| Di vs Monitor, unweighted | 4.75 | 16.25 | 99.59 | 68.57 | 95.58 |
| Reid vs Monitor, unweighted | 5.21 | 46.23 | 97.45 | 49.95 | 97.06 |



\end{document}